\documentclass[aps,twocolumn,nofootinbib,prl]{revtex4}

\usepackage{graphicx}
\usepackage{epic}
\usepackage{eepic}
\usepackage{latexsym}

\newcommand{\eq}[1]{(\ref{#1})}
\newcommand{\be}{\begin{equation}}
\newcommand{\ee}{\end{equation}}
\newcommand{\bea}{\begin{eqnarray}}
\newcommand{\eea}{\end{eqnarray}}

\newcommand{\vs}[1]{\vspace{#1 mm}}
\newcommand{\hs}[1]{\hspace{#1 mm}}

\def\b{\beta}

\def\d{\delta}

\def\f{\phi}
\def\fr{\frac}

\def\l{\lambda}

\def\m{\mu}
\def\n{\nu}

\def\r{\rho}
\def\s{\sigma}
\def\S{\Sigma}

\def\O{\Omega}

\def\o{\omega}

\def\del{\partial}

\def\nn{\nonumber}

\begin{document}

\title{D-brane Gases and 
Stabilization of Extra Dimensions in Dilaton Gravity}

\author{Savas Arapoglu}
\email[]{arapoglu@boun.edu.tr}
\affiliation{Department of Physics, Bo\~{g}azi\c{c}i University,\\
Bebek, 34342, \.Istanbul, Turkey} 

\author{Ali Kaya}
\email[]{kaya@gursey.gov.tr}
\affiliation{Feza G\"{u}rsey Institute,\\
\c{C}engelk\"{o}y, 81220, \.Istanbul, Turkey\vs{3}} 

\date{\today}

\begin{abstract}
We consider a toy cosmological model with a gas of wrapped $Dp$-branes
in 10-dimensional dilaton gravity compactified on a $p$-dimensional
Ricci flat internal manifold. A consistent generalization of the
low energy effective field equations in the presence of a conserved brane
source coupled to dilaton is obtained. It is then shown that the
compact dimensions are 
dynamically stabilized in string frame as a result of a balance
between negative winding and positive momentum pressures. Curiously,
when $p=6$, i.e. when the observed space is three dimensional, the
dilaton becomes a constant and stabilization in Einstein frame is also
realized.  
\end{abstract}

\maketitle

One of the main problems of string cosmology is to determine why the
extra compact dimensions evolve differently from the observed three
dimensions and remained comparatively very small. In \cite{bv}, an
intuitive mechanism was proposed to accommodate this difference where
strings winding extra dimensions fall out of thermal equilibrium and
stop the cosmological expansion. In \cite{wb,wb2,pb,bw}, the arguments
of \cite{bv} are quantified by demonstrating stabilization in Einstein
and dilaton gravities using the energy momentum tensor for string
winding and momentum modes on the torus (see also 
\cite{ak,bg3,bg4,bg5,bg6,bg7,bg8,bg9,bg10,bg11,bg12,bg13,bg14,bg15,bg16,
  bg17} for recent work on brane gas cosmology).

As pointed out in \cite{ak}, strings may not be able to stabilize extra
dimensions when topology is different than a torus. Considering,
for example, a spherical internal space, there exists no stable
winding strings in the spectrum. One would then wonder if the higher
dimensional branes can play a role in these compactifications. 
In \cite{ak}, it is shown that a gas
of $p$-branes wrapping over a $p$-dimensional compact, Ricci flat,
internal manifold can stabilize the volume modulus. As for strings, it
turns out that there is a balance between the winding and the
vibrational momentum modes. The results of \cite{ak} are obtained in
Einstein gravity and the purpose of this letter is to generalize the
framework to dilaton gravity. Here, it is natural to consider a toy
model with a gas of winding $D$-branes. Since dilaton couples to $D$-brane
world-volume, it should be activated in this scenario.   

Our first aim is to generalize the equations of motion in the presence
of a {\it conserved} source coupled to dilaton. In string frame, the
low energy effective action (we set $H_{\m\n\r}=0$) is given by    
\be
S=\int d^{10}x\,\sqrt{-g}\,\,
e^{-2\phi}\,\,\left[R\,+\,4(\nabla\phi)^2\,+\,e^{a\f}\,
{\cal L}_m\right],\label{str} 
\ee
where we add a term to include the effects of  matter governed by the
Lagrangian ${\cal L}_m$. From this action, the field equations can be
found as  
\bea
R_{\m\n}&+&2\nabla_\m\nabla_\n\f\nn\\
&-&\fr{1}{2}\left[R+4\nabla^2\f-4
  (\nabla\f)^2\right]g_{\m\n}=e^{a\f}\,T_{\m\n},\label{f1}\\ 
R&+&4\nabla^2\f-4 (\nabla\f)^2=F,\label{f2}
\eea
where $T_{\m\n}$ is the matter energy momentum tensor
\be
T_{\m\n}=-\fr{1}{\sqrt{-g}}\,\fr{\del}{\del
  g^{\m\n}} \left(\sqrt{-g}\,{\cal L}_m\right)
\ee
and the extra contribution $F$ in \eq{f2} arises from the coupling of 
dilaton to ${\cal L}_m$.  

The conservation formula $\nabla_\m
T^{\m\n}=0$ plays the role of matter field equations. To determine
the unknown function $F$ we note the well known fact that
\eq{f2} can be viewed as a consequence of \eq{f1} and the contracted
Bianchi identity. Taking the divergence of \eq{f1} with $\nabla^\m$,
one finds: 
\be
(\nabla_\n\f)\,F=(a-2)\,e^{a\f}\,T_{\n\l}\,\nabla^\l\f.\label{c}
\ee
Although in general $F$ cannot be uniquely fixed by this constraint,
in a cosmological context one can assume that 
\be
\f=\f(t),\hs{4}g_{ti}=0,\hs{4}T_{ti}=0,
\ee
where $t$ is the time coordinate and $i$ denotes a spatial
direction. Under these conditions, \eq{c} gives
\be
F=(a-2)\,e^{a \f}\,T_{tt}\,g^{tt}=-(a-2)\,e^{a \f}\,\r.\label{ff}
\ee
It is easy to see that \eq{ff} corresponds to the choice  
\be
{\cal L}_m\,=\,-\r\, ,
\ee
which is precisely the Lagrangian for hydrodynamical matter (see
e.g. section 10.2 of \cite{rev}). 

Let us take a $(1+m+p)$-dimensional space-time with the metric   
\be
ds^2=-dt^2+dx^idx^i+R^2\, d\S_p^2 \label{stmet},
\ee
and consider a $p$-brane wrapping over $\S_p$. Here $i,j=1,..,m$,
$(a,b =1,..,p)$, $\S_p$ is a $p$-dimensional compact manifold and $R$
is a constant scale factor. The brane dynamics is governed by the
usual action   
\be\label{ba}
S_p\,=\,-T_p\,\int\, d^{p+1}\s\,\sqrt{-\gamma},
\ee
where $T_p$ is the tension, $\gamma$ is the induced metric and
the local brane coordinates are identified as $\s=(t,\S_p)$. It is
easy to see that the wrapped brane offers a stable classical configuration
which solves the embedding equations derived from \eq{ba}. Although it
is not known how to quantize \eq{ba} exactly, for our purposes it
is enough to employ an approximate scheme and simply consider small
fluctuations around the classical background. It is clear that in this
theory there are both winding and vibrational momentum modes in the
spectrum. The energy of a winding mode is given by    
\be
E \,=\,(n\,T_p \,\O_p)\,R^p, \label{w}
\ee
where $n$ is the winding number and $\O_p$ is the volume of
$\S_p$. The spectrum of small vibrations can be found by expanding the
action \eq{ba} to the second order in transverse fluctuation fields
which yields free modes propagating in $\S_p$. The energy spectrum is
given by \cite{ak}   
\be
E\,=\,\fr{\l_k}{R} \label{m}, 
\ee 
where $-\l_k^2$ are the discrete eigenvalues of the Laplacian on
$\S_p$. For $D$-branes, there are additional $U(1)$ gauge fields
living on the world-volume. At the linearized level, the transverse
excitations obey free field equations and the energy spectrum is
identical to \eq{m}.  

The {\it average} pressure of a closed system can be found from the
energy as a function of volume; $E(V)$. Recalling that in our case the
volume modulus is fixed by $R^p$ one gets   
\be
P=\left\{\begin{array}{ll} 
-(n\,T_p \,\O_p)\,R^p:\,\,\,\textrm{winding},\\
\,\,\,\,\,\,\l_k/(pR)\hs{5}:\,\,\,\textrm{momentum}.\end{array}\right.
\label{wmp}
\ee
Note that the winding and the momentum modes apply negative and
positive pressures, respectively. For momentum modes, $1/R$ dependence
of the energy and the pressure can also be inferred from the
uncertainty principle 
and the fact that the size of a compact direction is approximately
equal to $R$.     
  
For a non-interacting gas of such excitations, it is enough to sum
over the modes with constant number densities to find the
total energy and pressure. However, it is more natural to assume a
state of thermal equilibrium\footnote{As discussed in \cite{pb}, 
  gravitons and photons offer an ideal candidate for a thermal bath
  which does not affect the dynamics of extra dimensions. In \cite{ak}
  we also observe that brane winding and momentum modes dominate the
  cosmology at late times in the presence of matter which has equation
  of state $p=\o\r$ with $\o>0$. Therefore, the contribution of the
  thermal bath can be ignored in the following.}
and consequently allow transitions between different energy levels. In
that case, the mean number of a mode in the gas is proportional to the
Boltzmann factor $e^{-\b E}$. At very low temperatures the gas mainly
consists of the lowest energy modes since the relative densities of
other excitations are exponentially suppressed. In this case, one can
use the expressions \eq{w}-\eq{wmp} with fixed quantum numbers
(corresponding to lowest lying states) to obtain the average energy
and pressure of the gas.    

In a cosmological setting, one should replace \eq{stmet} with 
\be\label{met1}
ds^2=-dt^2+e^{2B}dx^idx^i+e^{2C}\,d\S_p^2,
\ee
where the metric functions $B$ and $C$ depend only on the proper time
$t$. In this work we assume that $d\S_p^2$ is Ricci flat. The scale
factors for the observed and the internal spaces are defined by  
\be \label{robrin}
R_{ob}=e^{B},\hs{8}R_{in}=e^{C}.
\ee
To find the energy momentum tensor from \eq{w}-\eq{wmp}, 
the constant $R$ in these formulas should be replaced
with the time dependent function $e^{C}$ and one should also divide by
the total spatial volume $e^{mB+pC}$ to find the corresponding
densities. This gives  
\bea
T_{\hat{t}\hat{t}}&=&T_w\, e^{-mB}\,+\,T_m\,e^{-mB-(p+1)C},\nonumber\\
T_{\hat{i}\hat{j}}&=&0,\label{enmom1}\\
T_{\hat{a}\hat{b}}&=&-T_w \,e^{-mB}\,\d_{ab}\,+\,\fr{T_m}{p}\,
e^{-mB-(p+1)C}\,\d_{ab},\nonumber 
\eea
where  $T_w$ and $T_m$ are constants. It is easy to check that
$T_{\m\n}$ is conserved; $\nabla_\m T^{\m\n}=0$. As noted in
\cite{ak}, \eq{enmom1} implies an equation of state
$p_{\hat{i}}=\omega_{i}\,\rho$, where for winding modes  
\be\label{w1}
\omega_i=\left\{\begin{array}{ll} 
-1:\,\,\,\textrm{brane direction},\\
\,\,0\hs{2}:\,\,\,\textrm{transverse direction},\end{array}\right.
\ee
and for momentum modes
\be\label{w2}
\omega_i=\left\{\begin{array}{ll} 1/p:\,\,\,\textrm{brane direction},\\
\,\,\,\,0\,\,:\,\,\,\,\textrm{transverse direction},\end{array}\right.
\ee
which is equivalent to radiation confined in the compact space.

Up to now our considerations are general and the formulas are valid
for different brane types. To focus on $D$-branes we should fix the
value of $a$. Since $D$-branes contribute first at disk
order in string perturbation theory, their tension is inversely
proportional to string coupling $g_s=e^\f$. Therefore, one should
set\footnote{For fundamental strings the tension does not depend on
  $g_s$ and one should set $a=2$. In that case the field equations
  \eq{f1}-\eq{f2} agree with that of \cite{wb}.} 
\be 
a=1
\ee
to have the correct $g_s$ dependence in \eq{str}. Recall that we have
also   
\be
m+p=9
\ee
in string theory. 

Using \eq{enmom1} in \eq{f1}, \eq{f2} and \eq{ff} one can get the
following equations (we take some linear combinations to simplify the
expressions):   
\bea
&&\ddot{\f}=\,-k\,\dot{\f}+\fr{3-p}{2}\,F_1+ 2 \, F_2,\nn\\
&&\ddot{B}=\,-k\,\dot{B}+\fr{1}{2}\,F_1+ \fr{1}{2}
\,F_2,\label{de}\\ 
&&\ddot{C}=\,-k\,\dot{C}-\fr{1}{2}\, F_1+ \fr{p+2}{2p}\, F_2,\nn\\
&&k^2=m\,\dot{B}^2\,+\,p\,\dot{C}^2\,+\,2\,F_1\,+\,2\,F_2,\label{incon} 
\eea
where 
\bea
&&k\,=\,m\,\dot{B}+p\,\dot{C} - 2\,\dot{\f}, \nn\\
&&F_1\,=\,T_w\,e^{\f-mB},\label{frc}\\
&&F_2\,=\,T_m\,e^{\f-mB-(p+1)C},\nn
\eea
and dot denotes differentiation with respect to $t$. Eq. \eq{incon}
can be viewed as a constraint on initial data and when it is
obeyed at some $t_0$ it will be fulfilled during subsequent
evolution.  

We could not obtain the most general solution of these equations. Using,
however, the following ansatz 
\bea
&&\f\,=\, \f_1\,\ln (t)\,+\,\f_0,\nn\\
&&B\,=\, b_1\,\ln (t),\\
&&C\,=\, C_0,\nn
\eea
we find that \eq{de} gives
\bea
&&b_1=\fr{4}{m+3},\hs{4}\f_1=\fr{2\,(m-3)}{m+3},\\
&&e^{\f_0}=\fr{4\,(p+2)\,p}{T_w\,(p+1)(m+3)^2},\label{f0}\\
&&e^{(p+1)C_0}=\fr{(p+2)\,T_m}{p\,T_w},\label{c0} 
\eea
and \eq{incon} is identically satisfied. 

In summary, we have the following power-law solution:
\bea
&&ds^2=-dt^2\,+\,(t)^{8/(m+3)}\,dx^idx^i+e^{2C_0}\, d\S_p^2,\nn\\
&&e^\f=e^{\f_0}\,(t)^{2(m-3)/(m+3)}\, .\label{sol}
\eea
It is obvious that in this background the internal dimensions
are stabilized in string frame. Remarkably, when the observed
space is three dimensional (i.e. when $m=3$) the dilaton does not
depend on $t$ and the canonical Einstein frame becomes identical to
string frame. Again for $m=3$, the expansion of the observed space is
equivalent to the one for pressureless matter in standard
cosmology. When $m>3$ the dilaton grows and at some point we enter the
strong coupling regime in which the low energy field equations can no
longer be trusted.  

Apparently, \eq{sol} corresponds to special initial conditions and to
be able to talk about {\it dynamical} stabilization we should consider
more general initial data. To verify stabilization we did several numerical 
integrations with arbitrary initial conditions. Before presenting our
findings let us discuss the following issue. The constraint \eq{incon}
divides the solution space into two disjoint pieces according to the
sign of $k$ defined in \eq{frc}, since by \eq{incon} $k$ cannot vanish
and thus its sign cannot alter in time. The sign is strongly correlated with
the arrow of time in the problem. In all our numerical integrations we
observe that when $k<0$ a singularity at finite proper time is
encountered. These solutions can be thought to evolve backwards in
time and hit the initial ``big bang'' singularity in finite
duration. On the other hand all numerical integrations with $k>0$
yield smooth solutions and these can be thought to evolve forwards in
time.  

There is also a physically unexpected behavior for negative
$k$. From the second equation in \eq{de} we find that for $k<0$ and
when {\it initially} $\dot{B}>0$, $\ddot{B}$ is {\it always}
positive\footnote{It is easy to see that the analogous argument for
  $k>0$ does not work.} which implies an accelerated expansion for the
observed space since the scale factor consequently obeys
$\ddot{R}_{ob}>0$. This, however, contradicts with the anticipation 
that the winding and the momentum modes produce a deceleration
by slowing down the expansion since from \eq{w1} and \eq{w2} they do not
apply any pressure along the observed directions, i.e. they act like
pressureless dust. This shows that (at least some) solutions with
$k<0$ are not physically well-behaved.      

\begin{figure}
\centerline{\includegraphics[width=5.5cm]{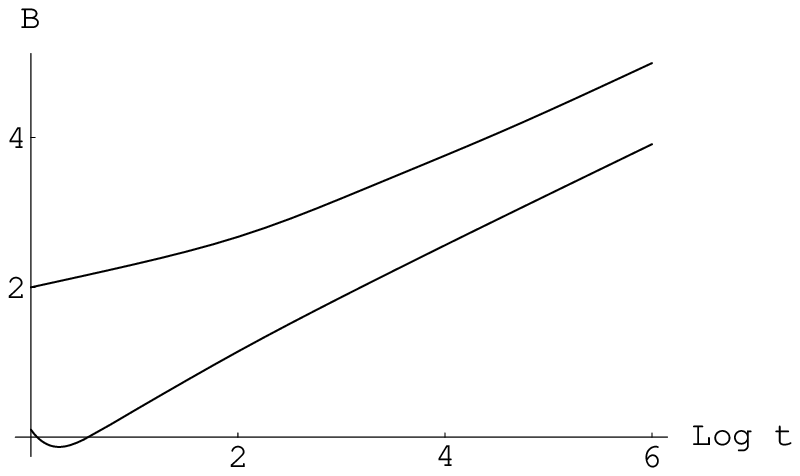}}
\centerline{\includegraphics[width=6.0cm]{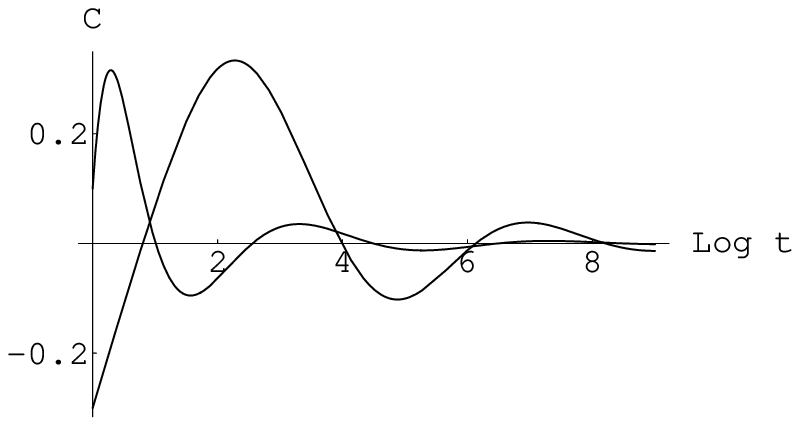}\,\,\,\,\,}
\centerline{\includegraphics[width=5.5cm]{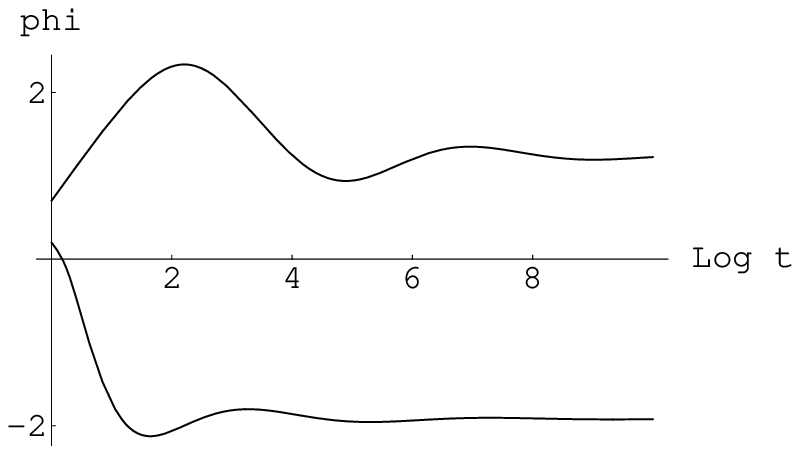}}
\caption{\label{fig1}$(m,p)=(3,6)$: The graphs of $(B,C,\f)$ for the
  following initial conditions 
  $B(1)=(0.1,2.0)$, $\dot{B}(1)=(-2.0,0.3)$, $C(1)=(0.1,-0.3)$,
  $\dot{C}(1)=(1.7,0.4)$ $\f(1)=(0.2,0.7)$, $\dot{\f}(1)=(-1.0,1.0)$,
  respectively. We choose 
  $T_w=4$ and $T_m=3$ so that the internal dimensions are stabilized at $C=0$.}
\end{figure}

\begin{figure}
\centerline{\includegraphics[width=5.5cm]{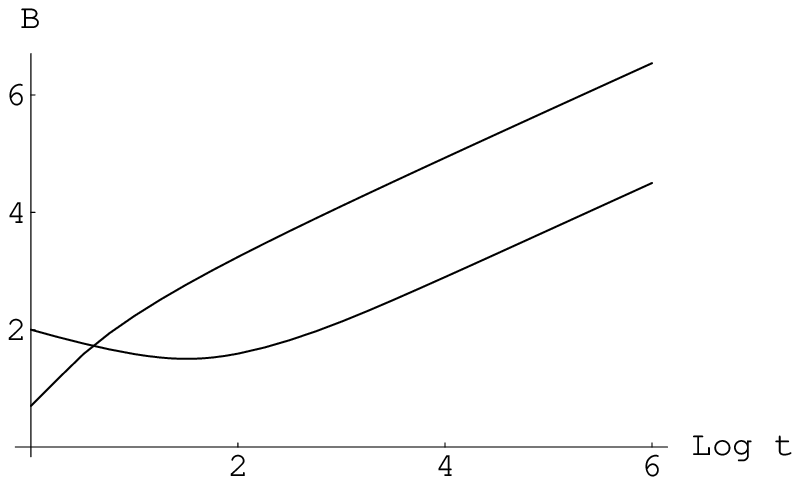}}
\centerline{\includegraphics[width=6.0cm]{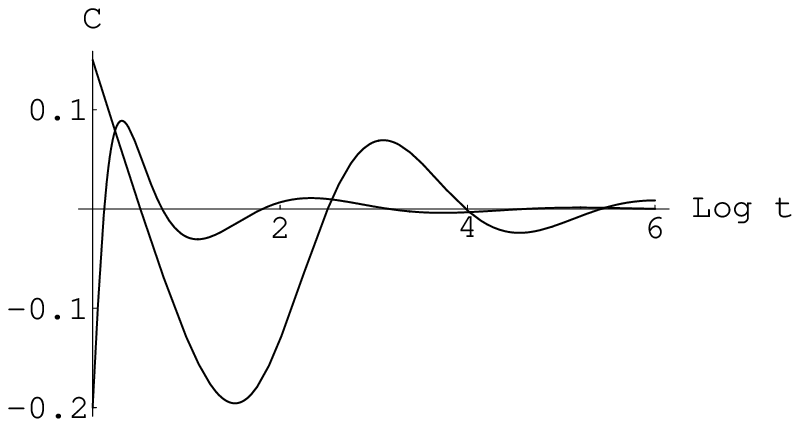}\,\,\,\,\,}
\centerline{\includegraphics[width=5.5cm]{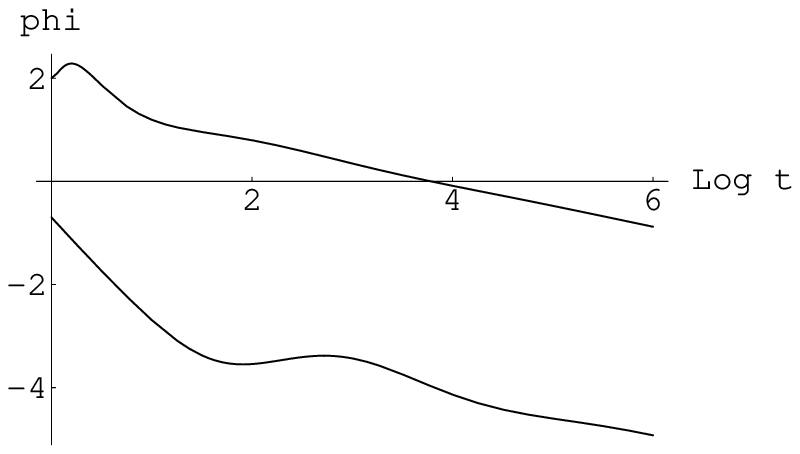}}
\caption{\label{fig2}$(m,p)=(2,7)$: The evolution of $(B,C,\f)$ for
  the initial data $B(1)=(0.7,2.0)$, $\dot{B}(1)=(0.9,-0.5)$,
  $C(1)=(-0.2,0.2)$, $\dot{C}(1)=(1.7,-0.3)$ $\f(1)=(2,-0.7)$,
  $\dot{\f}(1)=(0.3,-2.1)$, respectively. We set $T_w=9$, $T_m=7$ and
  as a result $C_0=0$ from \eq{c0}.} 
\end{figure}

\begin{figure}
\centerline{\includegraphics[width=5.5cm]{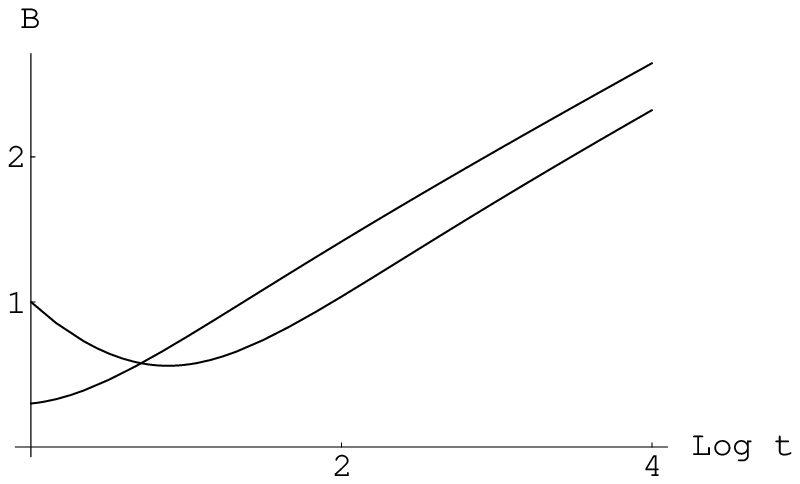}}
\centerline{\includegraphics[width=6.0cm]{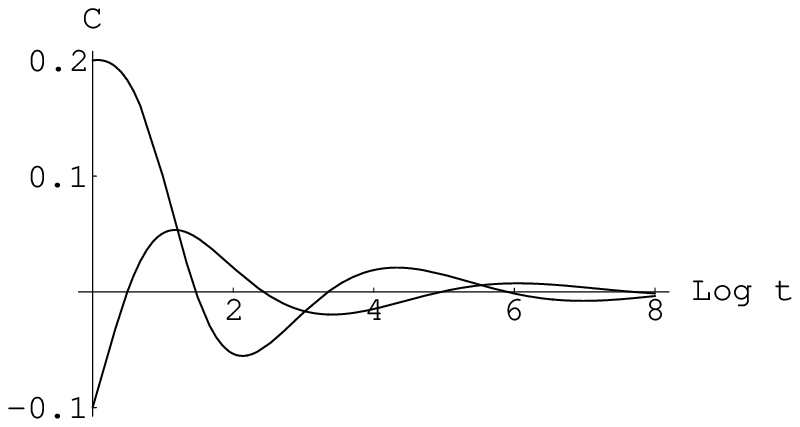}\,\,\,\,\,}
\centerline{\includegraphics[width=5.5cm]{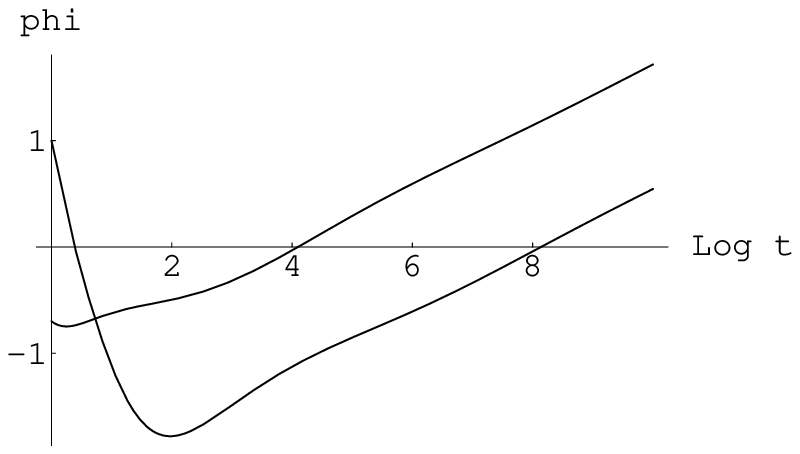}}
\caption{\label{fig3}
  $(m,p)=(4,5)$: The results of numerical integrations with the
  initial conditions 
  $B(1)=(1.0,0.3)$, $\dot{B}(1)=(-1.0,0.1)$, $C(1)=(0.2,-0.1)$,
  $\dot{C}(1)=(0.0,0.2)$, $\f(1)=(1,-0.7)$, $\dot{\f}(1)=(-3.0,-0.4)$,
  respectively. We have also $T_w=7$ and $T_m=5$.} 
\end{figure}

Focusing, therefore, on initial data with $k>0$, we observe that in
all numerical integrations the metric and the dilaton asymptotically
approach \eq{sol}. In figures \ref{fig1}, \ref{fig2} and \ref{fig3} we
plot two illustrative examples for $m=3,2,4$, respectively.  In these
runs while $B$ eventually increases in time, $C$ performs damped
oscillations around the constant value \eq{c0}. For $m>3$ the dilaton
grows, for $m<3$ it decreases and for $m=3$ it also oscillates around
\eq{f0} which are damped in time. These results strongly indicate that
\eq{sol} is a future asymptotic ``fixed point'' for \eq{de} and
\eq{incon}.   


To support these numerical findings analytically let us note that
\eq{de} can be viewed to describe a point particle in $(\f,B,C)$
space. There are three forces acting on the particle: two of them are
characterized by the functions $F_{1,2}$ and there is also a {\it
  frictional force} which is determined by the time dependent
coefficient $k$. The fact that we restrict $k>0$ is crucial in
identifying the frictional force\footnote{For $k<0$ there is an
  ``anti-frictional'' force. Not surprisingly this would destabilize
  the system and create a singularity. The existence of friction
  rather than anti-friction also signals that flow of time in the
  problem is correctly identified.}. Ignoring friction for a moment,
the third equation in \eq{de} shows that the driving force along $C$ vanishes
at $C=C_0$, where $C_0$ is given by \eq{c0}. $C=C_0$ is a stable
equilibrium point since when $C>C_0$ the net force is negative and
when $C<C_0$ it is positive. For stabilization it is also crucial that
the oscillations around $C_0$ are {\it damped} and this is exactly 
achieved by friction. Therefore, as $t\to\infty$
one expects $C\to C_0$ which proves stabilization.  

On the other hand, using the asymptotic value $C_0$ in \eq{de} one
sees that the sum of the forces generated by $F_{1,2}$ in $\f$
direction only vanishes for $m=3$ ($p=6$). In this case, due to the
existence of friction the motion in $\f$ should eventually stop and
this explains the stabilization of dilaton when $m=3$. For $m\not=3$,
the net force never vanishes and thus $\f$ will be time
dependent. Similarly, the second equation in \eq{de} shows that the
net force along $B$ direction is never zero and thus $B$ cannot
become a constant.   

Technically, it is necessary for stabilization that $F_1$ and $F_2$
terms in the third equation in \eq{de} have opposite signs. The value
of $a$ is crucial for this condition to hold. In general the equation
for $C$ takes the form    
\be
\ddot{C}=\,-k\,\dot{C}-\fr{a}{2}\, F_1+ \fr{2+(2-a)p}{2p}\, F_2.
\ee
For $a\leq0$ both terms become positive and stabilization cannot
be achieved. In the interval $0<a\leq 2$ the forces have opposite signs and 
for $a>(2/p)+2$ the stabilization is again ruined. Remarkably for
$D$-branes $a=1$, which is in the range that allows stabilization. 

Concluding, in this paper we have presented a dynamical stabilization
mechanism for the volume modulus in string theory based on the
cosmological impact of a gas of $D$-branes. The model is obviously
incomplete, however it reveals a generic dynamical behavior. It is not
clear if this mechanism can be used in a realistic scenario to ensure
moduli stabilization. One problem is that a straightforward
generalization of Brandenberger and Vafa (BV) mechanism \cite{bg3}
implies all $p$-branes with $p>2$ to annihilate in the early
universe. Therefore, having a gas of higher dimensional branes at late
times seems unlikely. On the other hand, as emphasized before, it
seems that higher dimensional branes are needed for moduli
stabilization in compactifications on topologically non-toroidal
spaces like spheres or phenomenologically viable Calabi-Yau or G2
manifolds. Therefore, to have a realistic scenario involving brane
gases it seems that we should either modify BV mechanism or improve
the method based on winding strings to circumvent the topological
restrictions.

\end{document}